\documentclass[pra,showpacs,twocolumn]{revtex4}
\usepackage{amsmath,amssymb,amsfonts,graphicx}

\renewcommand{\vec}[1]{\mathbf{#1}}
\newcommand{\bra}[1]{\mbox{$\langle#1|$}}
\newcommand{\ket}[1]{\mbox{$|#1\rangle$}}
\newcommand{\sprod}[2]{\mbox{$\langle#1|#2\rangle$}}
\newcommand{\kket}[1]{\mbox{$|#1\rangle$}}
\newcommand{\vv}{\textsf{v}}
\newcommand{\rep}[1]{\textsf{#1}}
\newcommand{\repp}[1]{\mathbb{#1}}
\newcommand{\ind}[1]{{\textsf{#1}}}
\newcommand{\reg}[1]{\mathcal{#1}}

\begin{document}

\title{Einstein--Podolsky--Rosen correlations of Dirac particles --
  quantum field theory approach}  

\author{Pawe{\l} Caban}\email{P.Caban@merlin.fic.uni.lodz.pl}
\author{Jakub
  Rembieli\'nski}\email{J.Rembielinski@merlin.fic.uni.lodz.pl}
\affiliation{Department of Theoretical Physics,
University of Lodz\\ 
Pomorska 149/153, 90-236 {\L}\'od\'z, Poland}

\date{\today}

\begin{abstract}
We calculate correlation function in the Einstein--Podolsky--Rosen
type of experiment with massive relativistic Dirac particles in the
framework of the quantum field theory formalism. We perform our
calculations for states which are physically interesting and
transforms covariantly under the full 
Lorentz group action, i.e.\ for pseudoscalar and vector
state. 
\end{abstract}


\maketitle

\section{Introduction}

One of the most puzzling aspects of quantum mechanics, its
nonlocality, is illustrated by the Einstein--Podolsky--Rosen (EPR)
paradox \cite{cab_EPR1935}. Quantum mechanics predicts that for a pair
of entangled particles, flying apart from each other, measurements
give results 
incompatible with our intuitive conceptions about reality and
locality. Quantum mechanical predictions have been confirmed in many
EPR--type experiments. Most of these experiments have been performed
with photon pairs. However all EPR experiments with photons are
subject to the so called detection loophole---low efficiency of photon
detection allows the possibility that the subensemble of detected
events agrees with quantum mechanics even though the entire ensemble
satisfies the requirements of local realism. Therefore the
fair--sampling hypothesis, stating that the detected events fairly
represent the entire ensemble, must be assumed. Detection loophole has
been closed recently in the experiment with massive particles
\cite{cab_RKMSIMW2001}. This experiment in turn does not overcome the
so called locality loophole in which the correlations of apparently
separate events could result from unknown subluminal signals
propagating between different particles. (Locality loophole has been
closed in the experiment with photons \cite{cab_Weihs1998}.)
One can hope that future experiments with relativistic massive
particles could close both mentioned above loopholes.
Therefore such experiments seem to be
very interesting for the basics of quantum mechanics.

On the other hand in the last decade,
starting from Czachor's papers
\cite{Czachor1997_1,Czachor1997_2}, EPR correlations in the
relativistic context have been widely discussed \cite{ALH2002,ALHK2003,ALMH2003,AM2002,BT2005,CR2003_Wigner,%
CR2005,CW2003,Czachor2005,GA2002,GBA2003,GKM2004,Harshman2005,JSS2005,JSS2006,%
KS2005,KM2003,LMS2005,LY2004,LD2003,LD2004,LPT2003,MAH2003,PST2002,PT2003_2,%
PT2003_3,PT2004_1,PST2005,PS2003,RS2002,SL2004,TU2003_1,TU2003_2,Terno2003,%
Terno2005,TB2002,YWYNMX2004,ZBGT2001}. Unfortunately the 
incompleteness of the relativistic quantum mechanics formalism (e.q.\
lack of the covariant notion of localization)\footnote{at least if we
  accept standard clock 
  synchronization convention in special relativity. See in this
  context \protect\cite{cab_CR1999} where the consistent
  relativistic  quantum mechanics in the 
  framework of nonstandard synchronization scheme for clocks was
  formulated and \protect\cite{RS2002} where quantum correlations in
  this framework were discussed} 
causes that our understanding of relativistic aspects of quantum
information theory is far from being satisfactory. 
Moreover it is unclear which spin operator should be used in the
relativistic context (spin is not a self-contained,
irreducible geometrical object in the relativistic quantum
mechanics). There are 
several different operators which has 
the proper nonrelativistic limit (for the discussion see
Sec.~\ref{sec:RSO}). Measurement of quantum spin correlations in EPR
experiments could help us to decide which spin operator is more
appropriate. In the
present paper we use quantum field theory methods to calculate
correlation function for massive Dirac particle--anti--particle pair
in the EPR type experiment using particular, in our opinion the most
adequate, spin operator. Our results can be useful for the discussion
of EPR experiments in which the spin correlation function of
elementary particles is measured. Such an experiment has been
performed in seventies \cite{cab_LRM1976} but with the 
nonrelativistic particles (protons).

Entangled pairs of massive particles are usually created in the decay
processes of elementary particles (e.g.\ $\pi^0\to e^+e^-$, $Z^0\to e^+e^-$). 
Decaying particle has of course well defined
Poincar\'e--covariant state (e.g.\ $\pi^0$ is a pseudoscalar particle,
$Z^0$ four--vector one). Dynamics of the decay process is Poincar\'e
invariant therefore the two-particle entangled state of decay products
also posses well defined transformation properties with respect to the
full Poincar\'e group (compare \cite{Harshman2005}). In our paper we
classify such states and 
calculate correlation functions in the pseudoscalar and four--vector
states which correspond to the states of the pair created in the
$\pi^0$ and $Z^0$ decay, respectively.  
Our results are valid 
for any Dirac particle--anti--particle pair but we concentrate our
discussion on the $\pi^0\to e^+e^-$ and $Z^0\to e^+e^-$ decays because
particles produced in these decays are ultrarelativistic.

In Sec.~\ref{sec:QDF} we
establish notation and
briefly remind basic facts concerning free quantum Dirac field. In
Sec.~\ref{sec:TPS} we consider two--particle states and classify them
according to transformation properties with respect to the full
Lorentz group. Next section we devote to the discussion of the 
spin operator. Finally, in Sec.~\ref{sec:CF} we calculate explicitly
correlation function for the particle--anti--particle pair in the
pseudoscalar and vector state. The last section contains our
concluding remarks.

\section{The setting}
\label{sec:QDF}

Let the field operator $\Hat{\Psi}(x)$ fulfills the Dirac
equation 
 \begin{equation}
 (i \gamma^\mu \partial_\mu - m)\Hat{\Psi}(x) = 0,
 \label{cab:Dirac_equation_field}
 \end{equation}
where $\gamma^\mu$ are Dirac matrices (their explicit form used in the
present paper and related conventions can be found in the Appendix
\ref{app:Dirac_matrices}). The field 
transforms under Lorentz transformations according to: 
 \begin{equation}
 U(\Lambda) \Hat{\Psi}(x) U(\Lambda^{-1}) = \textsf{D} (\Lambda^{-1})
 \Hat{\Psi}(\Lambda x),  
 \label{cab:field_operator_transformation}
 \end{equation}
where $U(\Lambda)$ belongs to the unitary irreducible representation of the
Poincar\'e group and $\rep{D}(\Lambda)$ is the bispinor representation
of the Lorentz group (see Appendix \ref{cab:app_bispinor_rep}). Field
operator has the standard momentum expansion 
 \begin{multline}
 \Hat{\Psi}_\alpha(x) = (2\pi)^{-3/2}
 \sum\limits_{\sigma=\pm\frac{1}{2}} \int \frac{d^3\vec{k}}{2k^0} 
 \Big[ e^{-ikx}  u_{\alpha \sigma}(k) a_{\sigma}(k) \\
 +  e^{ikx} v_{\alpha\sigma}(k) b_{\sigma}^{\dag}(k) \Big],
 \label{cab:field_operator}
 \end{multline}
where $a_{\sigma}^{\dag}(k)$ ($b_{\sigma}^{\dag}(k)$) are creation
operators of the particle (antiparticle) with four-momentum $k$ and
spin component along $z$-axis equal to $\sigma$, and
$k^0=\sqrt{m^2+\vec{k}}$.  
These operators fulfill the standard canonical anticommutation
relations 
 \begin{gather}
 \{ a_{\sigma^\prime}(k^\prime),a^{\dag}_{\sigma}(k) \} = 2 k^0
 \delta^3(\vec{k}- \vec{k}^\prime) \delta_{\sigma\sigma^\prime},
 \label{cab:commutators_eq1}\\
 \{ b_{\sigma^\prime}(k^\prime),b^{\dag}_{\sigma}(k) \} = 2 k^0
 \delta^3(\vec{k}- \vec{k}^\prime) \delta_{\sigma\sigma^\prime}
 \label{cab:commutators_eq2}
 \end{gather}
and all other anticommutators vanish. The
one-particle and antiparticle state with momentum $k$
and spin component $\sigma$ are defined as 
 \begin{equation}
 \ket{k,\sigma}_a \equiv a_{\sigma}^{\dag}(k) \ket{0}, \quad 
 \ket{k,\sigma}_b \equiv b_{\sigma}^{\dag}(k) \ket{0}, 
 \label{cab:one_particle_states}
 \end{equation}
respectively.
Here $\ket{0}$ denotes Lorentz--invariant vacuum, $\sprod{0}{0}=1$,
$a_\sigma(k) \ket{0} = 0$.
The states (\ref{cab:one_particle_states}) span the carrier
space of the irreducible unitary 
representation $U(\Lambda)$ of the Poincar\'e group
 \begin{equation}
 U(\Lambda) \ket{k,\sigma}_{a/b} =
 \mathcal{D}(R(\Lambda,k))_{\lambda\sigma} 
 \ket{\Lambda k,\lambda}_{a/b},
 \label{cab:one_particle_transf}
 \end{equation}
where ${\mathcal{D}}\in\text{SU(2)}$ is the matrix spin $1/2$
representation of the $SO(3)$ 
group, $R(\Lambda,k)=L_{\Lambda k}^{-1}\Lambda L_k$ is the Wigner
rotation and $L_k$ denotes the standard Lorentz boost defined by
the relations $L_k\tilde{k}=k$, $L_{\tilde{k}}=I$,
$\tilde{k}=(m,\vec{0})$. As follows from Eqs.\
(\ref{cab:commutators_eq1}--\ref{cab:commutators_eq2}), 
states (\ref{cab:one_particle_states}) are
normalized covariantly
 \begin{equation}
 {}_{a/b}\sprod{k,\sigma}{k^\prime,\sigma^\prime}_{a/b} = 2 k^0
 \delta^3(\vec{k}-\vec{k}^\prime) \delta_{\sigma\sigma^\prime}.
 \label{cab:normalization}
 \end{equation}
Eqs.\ (\ref{cab:field_operator_transformation},%
\ref{cab:one_particle_states},\ref{cab:one_particle_transf}) imply
standard consistency (Weinberg) conditions for amplitudes 
 \begin{align}
 & v(\Lambda k) = \textsf{D}(\Lambda) v(k)
 \mathcal{D}^T(R(\Lambda,k)), \label{cab:Weinberg_cond_1}\\
 & u(\Lambda k) = \textsf{D}(\Lambda) u(k)
 \mathcal{D}^\dag(R(\Lambda,k)), \label{cab:Weinberg_cond_2}
 \end{align}
where $v(k)$ and $u(k)$ denotes matrices $[v_{\mu\alpha}(k)]$ and
$[u_{\mu\alpha}(k)]$, respectively. Explicit form of amplitudes 
depends on the chosen representation of gamma matrices. Under the
choice given in Appendix \ref{app:Dirac_matrices} amplitudes
can be written as
 \begin{align}
 & u(k) = i \vv(k) \sigma_2, \\
 & v(k) = \gamma^5\vv(k), 
 \end{align}
where, for for the sake of convenience, we have introduced the matrix 
 \begin{equation}
 \vv(k) = \frac{1}{2\sqrt{1+\frac{k^0}{m}}}
 \left(\begin{array}{c}
 \left( I+ \tfrac{1}{m}k^\mu\sigma_\mu \right)\sigma_2 \\
 \left( I+ \tfrac{1}{m}{k^\pi}^\mu\sigma_\mu \right)\sigma_2
 \end{array} \right),
 \label{cab:matrix_vv}
 \end{equation}
with $k^\pi=(k^0,-\vec{k})$, and 
$\sigma_i$, $i=1,2,3$, designate the standard Pauli matrices, $\sigma_0=I$%
\footnote{Note that in our paper \protect\cite{CR2005} the matrix
  $\protect\vv(k)$   was denoted by 
  $v(k)$. In the present paper we use sans--serif font, since we want
  to preserve the standard notation for amplitudes in Dirac field
  expansion (\ref{cab:field_operator}).}. 

Action of the charge conjugation \textsf{C} and space inversion
\textsf{P} on the Dirac field has the form: 
 \begin{gather}
 \textsf{C} \Hat{\Psi}(x) \textsf{C}^\dag = \eta_C \mathcal{C}
 \Hat{\bar{\Psi}}^T(x),
 \label{cab:charge_conjug_ala} \\
 \textsf{P} \Hat{\Psi}(x) \textsf{P}^\dag = \eta_P \gamma^0
 \Hat{\Psi}(x^\pi), 
 \label{cab:parity_1}
 \end{gather}
where $\Hat{\bar{\Psi}} = {\Hat{\Psi}}^\dag \gamma^0$,
$|\eta_P|=|\eta_C|=1$ and $\mathcal{C}$ is the charge conjugation
matrix (\ref{cab:charge_conjug}).

Operators \textsf{P} and \textsf{C} act on creation operators as
follows: 
 \begin{align}
 & \textsf{P} a^{\dag}_{\sigma}(k) \textsf{P}^\dag = \eta_{P}^{*}
 a^{\dag}_{\sigma}(k^\pi), 
 && \textsf{P} b^{\dag}_{\sigma}(k) \textsf{P}^\dag = - \eta_{P}
 b^{\dag}_{\sigma}(k^\pi), 
 \label{cab:parity_2}\\
 & \textsf{C} a_{\sigma}(k) \textsf{C}^\dag = \eta_C b_{\sigma}(k), 
 && \textsf{C} b^{\dag}_{\sigma}(k) \textsf{C}^\dag = \eta_C
 a^{\dag}_{\sigma}(k). 
 \label{cab:charge_conjug_amplitudes}
 \end{align}
It is convenient to introduce
covariant one--particle and anti--particle states
 \begin{equation}
 \kket{\alpha,k}_{a/b} = \vv_{\alpha\sigma}(k) \ket{k,\sigma}_{a/b}.
 \label{cab:covariant_vectors}
 \end{equation}
Such states has the following transformation properties
 \begin{equation}
 U(\Lambda) \kket{\alpha,k}_{a/b} =
 \rep{D}(\Lambda^{-1})_{\alpha\beta} \kket{\beta,k}_{a/b},
 \end{equation} 
implied by Eqs.\
(\ref{cab:one_particle_transf},\ref{cab:Weinberg_cond}), and 
 \begin{align}%
 & \textsf{P} \kket{\alpha,k}_{a} = \eta_{P}^{*}
 \gamma^{0}_{\alpha\beta} \kket{\beta,k^\pi}_{a},  
 &&
 \textsf{C} \kket{\alpha,k}_{a} = \eta_{C}^{*} \kket{\alpha,k}_{b},
 \label{cab:parity_action_cov_1}\\
 & \textsf{P} \kket{\alpha,k}_{b} = - \eta_{P}
 \gamma^{0}_{\alpha\beta} \kket{\beta,k^\pi}_{b},
 \label{cab:parity_action_cov_2}
 &&
 \textsf{C} \kket{\alpha,k}_{b} = \eta_{C} \kket{\alpha,k}_{a}.
 \end{align}

\section{Two--particle states}
\label{sec:TPS}

The main goal of our paper is to calculate spin correlation function
of two Dirac particles. 
We calculate correlations in the states consisting of one
particle and one anti--particle. The space of such two--particle
states is spanned by the vectors
 \begin{equation}
 \ket{(k,\sigma)_b,(p,\lambda)_a} \equiv b^{\dag}_{\sigma}(k)
 a^{\dag}_{\lambda}(p) \ket{0}.  
 \end{equation}
To analyze transformation properties of the states it is convenient to
introduce the covariant basis, analogous to
(\ref{cab:covariant_vectors}), defined by  
 \begin{equation}
 \kket{(\alpha,k),(\beta,p)} = \vv(k)_{\alpha\sigma}
 \vv(p)_{\beta\lambda} \ket{(k,\sigma)_b,(p,\lambda)_a}.
 \label{cab:cov_two_particle}
 \end{equation}
Notice that for simplicity we have omitted indices $a$ and $b$ in the
state vector on the left hand--side of
(\ref{cab:cov_two_particle}). Hereafter we use the
convention that in the two--particle state vector left pair of indexes
refers to anti--particle, right pair to particle, respectively.
Using (\ref{cab:Weinberg_cond}) we have
 \begin{multline}
 U(\Lambda) \kket{(\alpha,k),(\beta,p)}  \\ 
 = \rep{D}(\Lambda^{-1})_{\alpha\alpha^\prime} 
 \rep{D}(\Lambda^{-1})_{\beta\beta^\prime} \kket{(\alpha^\prime,\Lambda
   k),(\beta^\prime,\Lambda p)}.
 \label{cab:cov_two_particle_transf}
 \end{multline}
Moreover, we can easily determine the action of the discrete
operations $\textsf{P}$, $\textsf{C}$ on the states
(\ref{cab:cov_two_particle}) 
 \begin{align}
 & \textsf{P} \kket{(\alpha,k),(\beta,p)} = -  
  \gamma^{0}_{\alpha\alpha^\prime} \gamma^{0}_{\beta\beta^\prime} 
  \kket{(\alpha^\prime,k^\pi),(\beta^\prime,p^\pi)},
 \label{cab:parity_action_two_particle}\\
 & \textsf{C} \kket{(\alpha,k),(\beta,p)} = - 
 \kket{(\beta,p),(\alpha,k)}.
 \end{align}
Particle--anti--particle pairs are usually created in the state in
which the total four-momentum is determined as for example in the decay
$\pi^0\to e^-e^+$ where the total four-momentum of the electron--positron
pair is equal to the four-momentum of the decaying $\pi^0$. (This decay
channel has very small but nonzero width \cite{cab_PDG2004}.) The most
general particle--anti--particle state with total four-momentum $q$ has
the form 
 \begin{multline}
 \ket{\varphi}_q = \int \frac{d^3\vec{k}}{2k^0}
 \frac{d^3\vec{p}}{2p^0}  \delta^4(q-(k+p)) \\
 \times \Big[\sum_A \varphi^A(k,p) \mathcal{C} \Gamma_A \Big]_{\alpha\beta}
 \kket{(\alpha,k),(\beta,p)},
 \label{cab:general_definite_momentum}
 \end{multline}
where $\varphi^A(k,p)$ are numerical functions, 
$\mathcal{C}$ is given by (\ref{cab:charge_conjug})
and $\Gamma_A$ forms a subset of matrices
 \begin{equation}
 I, \gamma^5, \gamma^\mu, \gamma^\mu \gamma^5, [\gamma^\mu,\gamma^\nu]
 \label{cab:Clifford_basis}
 \end{equation}
which transform covariantly under Lorentz transformations and form a
basis of the Clifford algebra generated by Dirac gamma matrices. 
Matrices $\Gamma_A$
transform with respect to the
index $A$ according to certain representation $\repp{D}(\Lambda)$ of
the Lorentz group
 \begin{equation}
 \textsf{D}(\Lambda) \Gamma_A \textsf{D}(\Lambda^{-1}) =
 \repp{D}_{AB}(\Lambda^{-1}) \Gamma_B.
 \label{cab:general_state_transf_2}
 \end{equation}
Note that we have inserted matrix $\mathcal{C}$ in Eq.\
(\ref{cab:general_definite_momentum}) because 
from (\ref{cab:charge_conjug_condition}) we have 
$\textsf{D}^T(\Lambda) \mathcal{C} = \mathcal{C}
\textsf{D}^{-1}(\Lambda)$.

Identification of the singlet or vector state is based on the
transformation properties of the state
(\ref{cab:general_definite_momentum}) under Lorentz transformations
and parity \textsf{P}. Using (\ref{cab:cov_two_particle_transf}) we
have 
 \begin{multline}
 U(\Lambda) \ket{\varphi}_q =
 \int  \frac{d^3\vec{k}}{2k^0} \frac{d^3\vec{p}}{2p^0}
 \delta^4(\Lambda q-(k+p)) \\
 \times \Big[ \sum_A \varphi^{\prime A}(k,p) \mathcal{C} \Gamma_A \Big]_{\alpha\beta}
 \kket{(\alpha,k),(\beta,p)},
 \end{multline}
where from Eq.\ (\ref{cab:general_state_transf_2}) we obtain
 \begin{equation}
 \varphi^{\prime A}(k,p) = \repp{D}_{BA}(\Lambda^{-1})
 \varphi^B(\Lambda^{-1}k,\Lambda^{-1}p). 
 \label{cab_transformation_density}
 \end{equation}
Under parity the elements of the Clifford algebra transform as follows 
 \begin{equation}
 \gamma^0 \Gamma_A \gamma^0 = \repp{P}_{AB} \Gamma_B,
 \label{cab:covariant_parity_two}
 \end{equation}
where $\repp{P}$ represents parity in the carrier space of the
representation $\repp{D}$; namely, the state transforming
according to (\ref{cab:general_state_transf_2}) and
(\ref{cab:covariant_parity_two}) will be called: scalar if
$\repp{D}(\Lambda) = I$, $\repp{P} = I$, $\{\Gamma_A\} = \{I\}$;
pseudoscalar if
$\repp{D}(\Lambda) = I$, $\repp{P} = -I$, $\{\Gamma_A \}=
\{\gamma^5\}$; four--vector if 
$\repp{D}(\Lambda) = \Lambda$, $\repp{P} =
\eta=\text{diag}(1,-1,-1,-1)$ , $\{\Gamma_A\} = \{\gamma^\mu\}$ and
pseudo--four--vector if   
$\repp{D}(\Lambda) = \Lambda$, $\repp{P} = -\eta$, $\{\Gamma_A\} =
\{\gamma^\mu \gamma^5\}$. 

In this paper we restrict ourselves only to the pseudoscalar and
four--vector case since such states can be identified with real
particles decaying into $e^+e^-$ pair (compare for example $\pi^0$ and
$Z^0$ and their corresponding decay channels
\cite{cab_PDG2004}).

\paragraph{Pseudoscalar}
The general pseudoscalar state reads
 \begin{align}
 \ket{\varphi}_{q}^{\ind{ps}}  
  & = \int \frac{d^3\vec{k}}{2k^0}
 \frac{d^3\vec{p}}{2p^0} \delta^4(q-k-p) \varphi(k,p) \nonumber\\ 
  & \phantom{\rule{1cm}{1mm}} \times (\mathcal{C}\gamma^5)_{\alpha\beta}
 \kket{(\alpha,k),(\beta,p)} \nonumber\\
 & = \int \frac{d^3\vec{k}}{2k^0}
 \frac{d^3\vec{p}}{2p^0} \delta^4(q-k-p)
 \varphi(k,p) \nonumber\\ 
 & \phantom{\rule{1cm}{1mm}} \times
 \Big(\vv^{T}(k)\mathcal{C}\gamma^5 \vv(p) \Big)_{\sigma\lambda}
 \ket{(k,\sigma),(p,\lambda)}, 
 \label{cab:pseudoscalar_choosen}
 \end{align}
where, according to Eq.\ (\ref{cab_transformation_density}) the
density function transforms under Lorentz transformations as follows: 
 \begin{equation}
 \varphi^\prime(\Lambda k,\Lambda p) = \varphi(k,p).
 \end{equation}

\paragraph{Four--vector}
The most general four--vector state has the following form
 \begin{align}
 \ket{\varphi}_{q}^{\ind{vec}} & = \int \frac{d^3\vec{k}}{2k^0}
 \frac{d^3\vec{p}}{2p^0} \delta^4(q-k-p)
 \varphi^{\mu}(k,p) \nonumber\\
 & \phantom{\rule{2cm}{1mm}} \times (\mathcal{C}\gamma_\mu)_{\alpha\beta} 
 \kket{(\alpha,k),(\beta,p)} \nonumber\\
  & = \int \frac{d^3\vec{k}}{2k^0}
 \frac{d^3\vec{p}}{2p^0} \delta^4(q-k-p)
 \varphi^{\mu}(k,p) \nonumber\\ 
 & \phantom{\rule{1cm}{1mm}} \times
 \Big(\vv^{T}(k)\mathcal{C}\gamma_\mu \vv(p) \Big)_{\sigma\lambda}
 \ket{(k,\sigma),(p,\lambda)}. 
 \label{cab:state_vector_chosen}
 \end{align}
One can easily check that
 \begin{equation}
 \big( p_\mu + k_\mu \big) \vv^{T}(k) \mathcal{C} \gamma^\mu
 \vv(p) = 0.
 \end{equation}
Therefore, in the integral (\ref{cab:state_vector_chosen}) the nonzero
contribution has only such $\varphi^{\mu}(k,p)$, which
fulfills the transversality condition
 \begin{equation}
 q_\mu \varphi^{\mu}(k,p) = \big(p_\mu + k_\mu \big) \varphi^{\mu}(k,p) = 0. 
 \label{cab:transversality_cond}
 \end{equation}

Notice that in the decay cases the functions
$\varphi(k,p)$, $\varphi^\mu(k,p)$ are related to the dynamics of the
decay.

\section{Relativistic spin operator}
\label{sec:RSO}

To calculate correlation function we have to introduce the spin
operator for a relativistic massive particle. In the discussion of the
relativistic Einstein--Podolsky--Rosen experiment  various authors use
different spin operators 
\cite{ALMH2003,Czachor1997_1,RS2002,LY2004,LD2003,TU2003_1,TU2003_2}.
However it seems \cite{cab_BLT1969,CR2005,Terno2003} that the best
candidate for the relativistic spin operator is
 \begin{equation}
 \Hat{\vec{S}}= \frac{1}{m}\left(\Hat{\vec{W}} -
   \Hat{W}^0\frac{\Hat{\vec{P}}}{\Hat{P}^0+m} \right), 
 \label{spin}
 \end{equation}
where $\Hat{W}^\mu$ is the Pauli--Lubanski four--vector
\begin{equation}
 \Hat{W}^\mu=\tfrac{1}{2}\varepsilon^{\mu\nu\sigma\lambda}\Hat{P}_\nu
 \Hat{J}_{\sigma\lambda}.
 \label{Pauli-Lubanski}
 \end{equation}
Here $\Hat{P}_\nu$ is a four-momentum operator and $\Hat{J}_{\sigma\lambda}$
denotes generators of the Lorentz group, i.e.,
$U(\Lambda)=\exp{(i\omega^{\mu\nu}\Hat{J}_{\mu\nu})}$. One
can show \cite{cab_BLT1969} that operator (\ref{spin}) is the only
operator which is a linear function of $W^\mu$ and fulfills the relations
\begin{subequations}
 \begin{gather}
 [\Hat{J}^i,\Hat{S}^j]=i\varepsilon_{ijk}\Hat{S}^k,\\
 [\Hat{S}^i,\Hat{S}^j]=i\varepsilon_{ijk}\Hat{S}^k,  
 \label{spin_commutator} \\ 
 [\Hat{P}^\mu,\Hat{S}^j]=0,\label{cab:commutator_spin_momentum}
 \end{gather}%
\label{spin_all_commutators}%
\end{subequations}%
and is a pseudovector, i.e.\ $\textsf{P} \Hat{S}^i \textsf{P} = \Hat{S}^i$.
%which should be satisfied for the spin operator.
Here $\Hat{J}^i=\frac{1}{2}\varepsilon_{ijk}\Hat{J}^{jk}$. 

In the representation (\ref{gamma_explicit}) of gamma matrices we have
 \begin{align}
 \Hat{W}^0 \kket{(\alpha,k)} & = W^{0}_{\alpha\beta}
 \kket{(\beta,k)}, \\
 \Hat{\Vec{W}} \kket{(\alpha,k)} & = \Vec{W}_{\alpha\beta} 
 \kket{(\beta,k)},
 \end{align}
where
 \begin{align}
 & W^0 = - \frac{1}{2} \begin{pmatrix} 
 \vec{k}{\cdot}\boldsymbol{\sigma} & 0 \\ 0 & 
 \vec{k}{\cdot} \boldsymbol{\sigma} 
 \end{pmatrix},\\
 & \vec{W} = - \frac{1}{2} \left[ 
 k^0 \boldsymbol{\sigma} \begin{pmatrix} I & 0 \\ 0 & I \end{pmatrix} 
 -i (\vec{k}\times \boldsymbol{\sigma}) 
 \begin{pmatrix} I & 0 \\ 0 & -I \end{pmatrix}
 \right].
 \end{align}

Therefore for the spin operator we easily find
 \begin{align}
 & \Hat{\Vec{S}} \ket{k,\sigma} = 
 \frac{1}{2} \boldsymbol{\sigma}_{\lambda\sigma}
 \ket{k,\lambda}, 
 \label{cab_spin_operator_action_standard}\\ 
 & \Hat{\Vec{S}} \kket{\alpha,k} = \left( \vv(k)
 \frac{\boldsymbol{\sigma}^T}{2}
 \bar{\vv}(k)\right)_{\alpha\beta} \kket{\beta,k}.  
 \end{align}
Real detectors register only particles which momenta belong to
some definite region in momentum space. Therefore, taking into account 
Eqs.\ %
(\ref{cab:commutator_spin_momentum},\ref{cab_spin_operator_action_standard}),
in one--particle subspace of the Fock space, the operator measuring spin of
the particle with 
four--momentum from the region $\reg{A}$ of the momentum space has the form
 \begin{equation}
 \Hat{\Vec{S}}^{a}_{\reg{A}} = \int_{\reg{A}}
 \frac{d^3\vec{k}}{2k^0}\, a^{\dag}_{\sigma}(k)
 \frac{\boldsymbol{\sigma}_{\sigma\lambda}}{2} a_{\lambda}(k).
 \label{cab:spin_creation_anih_particle}
 \end{equation}
This operator gives zero when acting on the anti--particle state
or state of the particle with four--momentum outside the region $\reg{A}$.
Analogous operator measuring spin of the anti--particle with
four--momentum belonging to the region $\reg{B}$ of the momentum space
can be written as
 \begin{equation}
 \Hat{\Vec{S}}^{b}_{\reg{B}} = \int_{\reg{B}}
 \frac{d^3\vec{k}}{2k^0}\, b^{\dag}_{\sigma}(k)
 \frac{\boldsymbol{\sigma}_{\sigma\lambda}}{2} b_{\lambda}(k). 
 \label{cab:spin_creation_anih_antiparticle}
 \end{equation}
We have
 \begin{equation}
 [\Hat{S}^{a,i}_{\reg{A}},\Hat{S}^{b,j}_{\reg{B}}] =0.
 \label{cab:spin_one_commutator}
 \end{equation}

\section{Correlation functions}
\label{sec:CF}

In this section we calculate correlation function in the
EPR--Bohm type experiment. In such an experiment we have two distant
observers, say Alice and Bob. We assume that both observers are at
rest with respect to the same inertial reference
frame. Particle--anti--particle pair is produced in an entangled state
$\ket{\varphi}$, the particle is registered by Alice while the
antiparticle by Bob. Alice
measures spin component of the particle along direction $\vec{a}$, Bob
spin component of the anti--particle along direction
$\vec{b}$. Therefore Alice uses the observable
$\vec{a}{\cdot}\Hat{\Vec{S}}^{a}_{\reg{A}}$ and Bob the observable $\vec{b}
{\cdot}\Hat{\Vec{S}}^{b}_{\reg{B}}$ (see Eqs.\
(\ref{cab:spin_creation_anih_particle},%
\ref{cab:spin_creation_anih_antiparticle})). So, by virtue of
(\ref{cab:spin_one_commutator}), the normalized correlation function 
in the state $\ket{\varphi}$ has the following form:
 \begin{equation}
 C_{\varphi}^{\reg{AB}}(\vec{a},\vec{b}) = 4\frac{ \bra{\varphi} \left(\vec{a}
     {\cdot}\Hat{\Vec{S}}^{a}_{\reg{A}} \right) \left(\vec{b}
     {\cdot}\Hat{\Vec{S}}^{b}_{\reg{B}} 
   \right) \ket{\varphi}}{\sprod{\varphi}{\varphi}}.
 \label{cab:correlation_general}
 \end{equation}
We calculate correlation function in two important cases when
EPR pair is produced in the pseudo--scalar or four--vector state.

\subsection{Pseudoscalar state}

The pseudoscalar state is given by Eq.\
(\ref{cab:pseudoscalar_choosen}). 
Therefore from (\ref{cab:correlation_general}) we find
\begin{widetext}
 \begin{multline}
 C_{\ind{ps}}^{\reg{AB}}(\vec{a},\vec{b}) = \\
 \Bigg\{\int \frac{d^3\vec{k}}{2k^0} \frac{d^3\vec{p}}{2p^0}
 \chi_{\reg{A}}(p) \chi_{\reg{B}}(k) \left(\delta^4(q-k-p)\right)^2 
 |\varphi(k,p)|^2 
 \text{Tr}\left[\vec{b}{\cdot}\boldsymbol{\sigma} \Big(\vv^{T}(k) \mathcal{C}
   \gamma^5 \vv(p) \Big) \vec{a}{\cdot}\boldsymbol{\sigma}^T 
  \Big(\vv^{T}(k) \mathcal{C} \gamma^5 \vv(p) \Big)^\dag
 \right]\Bigg\}  \\
 \times \Bigg\{\int \frac{d^3\vec{k}}{2k^0} \frac{d^3\vec{p}}{2p^0} 
 \chi_{\reg{A}}(p) \chi_{\reg{B}}(k) \left(\delta^4(q-k-p)\right)^2 
 |\varphi(k,p)|^2\,
 \text{Tr}\left[\Big(\vv^{T}(k) \mathcal{C} \gamma^5 \vv(p) \Big)  
  \Big(\vv^{T}(k) \mathcal{C} \gamma^5 \vv(p) \Big)^\dag
 \right]\Bigg\}^{-1} ,
 \label{cab:correl_function_pseudoscalar_general}
 \end{multline}
where $\chi_{\reg{A}}(p)$ i $\chi_{\reg{B}}(k)$ are characteristic
functions of the regions $\reg{A}$ and $\reg{B}$ in the corresponding
momentum spaces. Using Eqs.\ (\ref{cab:matrix_vv}, 
\ref{cab:charge_conjug}) and (\ref{gamma_explicit}) we arrive after little
algebra at 
 \begin{multline}
 \text{Tr}\Big[\vec{b}{\cdot}\boldsymbol{\sigma} \Big(\vv^{T}(k) \mathcal{C}
   \gamma^5 \vv(p) \Big) \vec{a}{\cdot}\boldsymbol{\sigma}^T 
  \Big(\vv^{T}(k) \mathcal{C} \gamma^5 \vv(p) \Big)^\dag
 \Big] = \\
 \frac{-1}{m^2}\Big\{\vec{a}{\cdot}\vec{b}(m^2+kp) - (\vec{k}\times
 \vec{p}) \Big[ \vec{a}\times\vec{b} +
 \frac{(\vec{b}{\cdot}\vec{p})(\vec{a}\times\vec{k}) - 
 (\vec{a}{\cdot}\vec{k})(\vec{b}\times\vec{p})}{(m+k^0)(m+p^0)}
  \Big]\Big\},
 \end{multline}
 \begin{equation}
 \text{Tr}\Big[\Big(\vv^{T}(k) \mathcal{C} \gamma^5 \vv(p) \Big)  
  \Big(\vv^{T}(k) \mathcal{C} \gamma^5 \vv(p) \Big)^\dag
 \Big] = \frac{m^2+kp}{m^2}.
 \end{equation}
Therefore in the simple situation when momenta of both particles in
the state $\ket{\varphi}^{\ind{ps}}_{q}$ are sharp (i.e.\ characteristic
functions in Eq.\ (\ref{cab:correl_function_pseudoscalar_general}) are
replaced by delta functions) we obtain the following correlation
function: 
 \begin{equation}
 C_{\ind{ps}}^{pk}(\vec{a},\vec{b}) = 
  -\vec{a}{\cdot}\vec{b} + 
 \frac{\vec{k}\times\vec{p}}{m^2+kp}
 \Big[ \vec{a}\times\vec{b} +
 \frac{(\vec{b}{\cdot}\vec{p})(\vec{a}\times\vec{k}) - 
 (\vec{a}{\cdot}\vec{k})(\vec{b}\times\vec{p})}{(m+k^0)(m+p^0)}
  \Big].
 \label{cab:correl_pseudoscalar_sharp}
 \end{equation}
In this special case of the sharp momenta we obtained the same
correlation function 
(\ref{cab:correl_pseudoscalar_sharp}) in our previous 
paper \cite{CR2005} where we discussed the Lorentz--covariant spin
density matrix in the framework of the relativistic quantum
mechanics. Notice that in the case when the laboratory frame (observers)
coincides with the center of mass 
frame, or even when $\vec{p}||\vec{k}$, from Eq.\
(\ref{cab:correl_pseudoscalar_sharp}) we get the same 
correlation function as for the singlet in the nonrelativistic case
 \begin{equation}
 C_{\ind{ps}}^{\ind{CMF}}(\vec{a},\vec{b}) = 
  -\vec{a}{\cdot}\vec{b}
 =C_{\ind{ps}}^{\vec{k}||\vec{p}}(\vec{a},\vec{b}) .
  \label{cab_correl_CMF_singlet}
 \end{equation}
It is interesting that correlation function calculated by Czachor in
\cite{Czachor1997_2}, in the center of mass frame still depends on
momentum and has the following form
 \begin{equation}
 C_{\ind{Czachor}}^{\ind{CMF}}(\vec{a},\vec{b}) = 
 - \frac{\vec{a}{\cdot}\vec{b} - \beta^2 \vec{a}_\perp{\cdot} \vec{b}_\perp}%
 {\sqrt{1+\beta^2[(\vec{n}{\cdot}\vec{a})^2-1]}
 \sqrt{1+\beta^2[(\vec{n}{\cdot}\vec{b})^2-1]}},
 \label{cab_correl_CMF_singlet_Czachor}
 \end{equation} 
where $\vec{n}=\vec{k}/|\vec{k}|$, $\beta=|\vec{k}|/k^0$,
$\vec{a}_\perp=\vec{a}-(\vec{n}{\cdot}\vec{a})\vec{n}$ and 
$\vec{b}_\perp=\vec{b}-(\vec{n}{\cdot}\vec{b})\vec{n}$.

Formulas (\ref{cab_correl_CMF_singlet}) and
(\ref{cab_correl_CMF_singlet_Czachor})   are different because Czachor
uses different spin operator. Therefore the experimental measurement
of the correlation function in the center of mass frame could show
which spin operator is more adequate in the relativistic quantum
mechanics. One of the possible sources of the electron--positron pairs
is the $\pi^0$ decay into the channel $\pi^0\to
e^+e^-$. Since $m_{\pi^0}=134.98 \; \text{MeV}/\text{c}^2$, 
$m_e=0.51 \; \text{MeV}/\text{c}^2$
\cite{cab_PDG2004}, electrons and positrons produced in this decay in
the center of mass frame are ultrarelativistic. Therefore let us find
the limit $\beta\to1$ of the formulas (\ref{cab_correl_CMF_singlet}) and
(\ref{cab_correl_CMF_singlet_Czachor}). We get
 \begin{gather}
 \left.C_{\ind{ps}}^{\ind{CMF}}(\vec{a},\vec{b}) \right|_{\beta\to1} = 
  -\vec{a}{\cdot}\vec{b},\\
 \left.C_{\ind{Czachor}}^{\ind{CMF}}(\vec{a},\vec{b})
 \right|_{\beta\to1} = 
 -\frac{(\vec{n}{\cdot}\vec{a})(\vec{n}{\cdot}\vec{b})}%
{|(\vec{n}{\cdot}\vec{a})(\vec{n}{\cdot}\vec{b})|} = \pm1, 
 \label{cab_correl_Czachor_limit_ultrarel}
 \end{gather}
respectively. We point out the discontinuity in 
the Czachor's correlation function in the ultrarelativistic limit
(\ref{cab_correl_Czachor_limit_ultrarel}).
It is also interesting to notice, that the function
 \begin{equation}
 \Delta C = C_{\ind{Czachor}}^{\ind{CMF}}(\vec{a},\vec{b})- 
 C_{\ind{ps}}^{\ind{CMF}}(\vec{a},\vec{b}) 
 \label{cab_correlation_difference}
 \end{equation}   
can take quite large values for $\beta$ close to 1. 
We shown this function in the Fig.\ \ref{rys} for the value
$\beta=0.999$ which corresponds to $e^+e^-$ created in the $\pi^0$
decay at rest.  

\begin{figure}
\rule{0mm}{1mm}\hfill
\includegraphics[width=0.45\columnwidth]{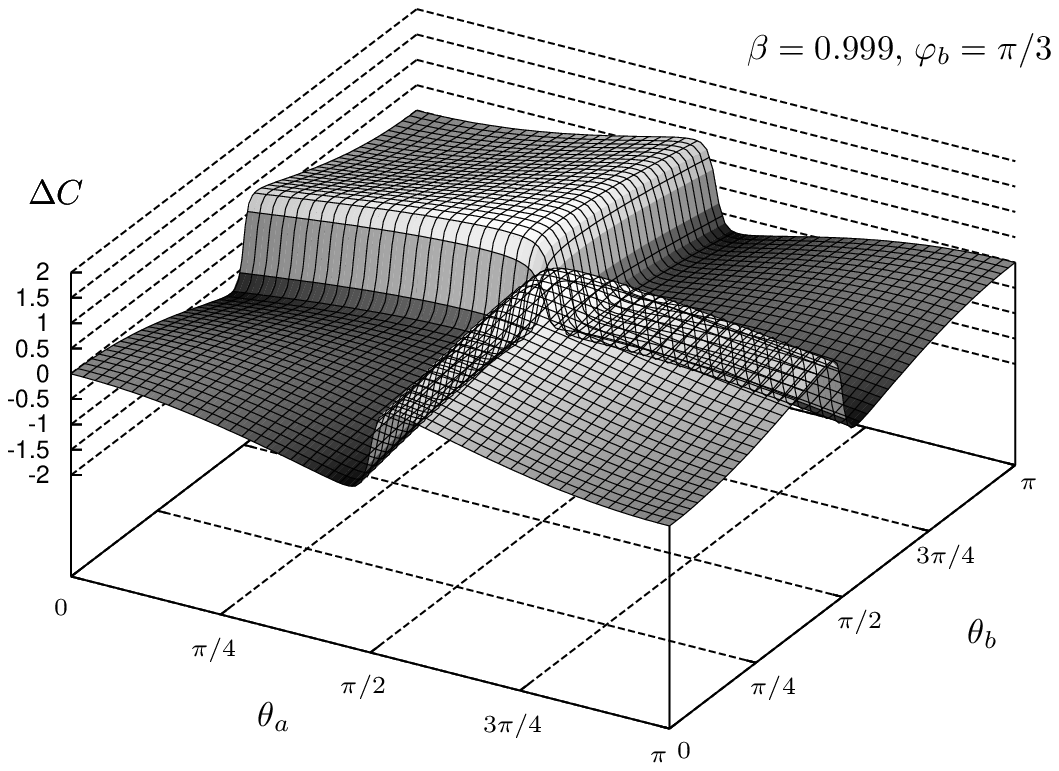}\hfill
\includegraphics[width=0.45\columnwidth]{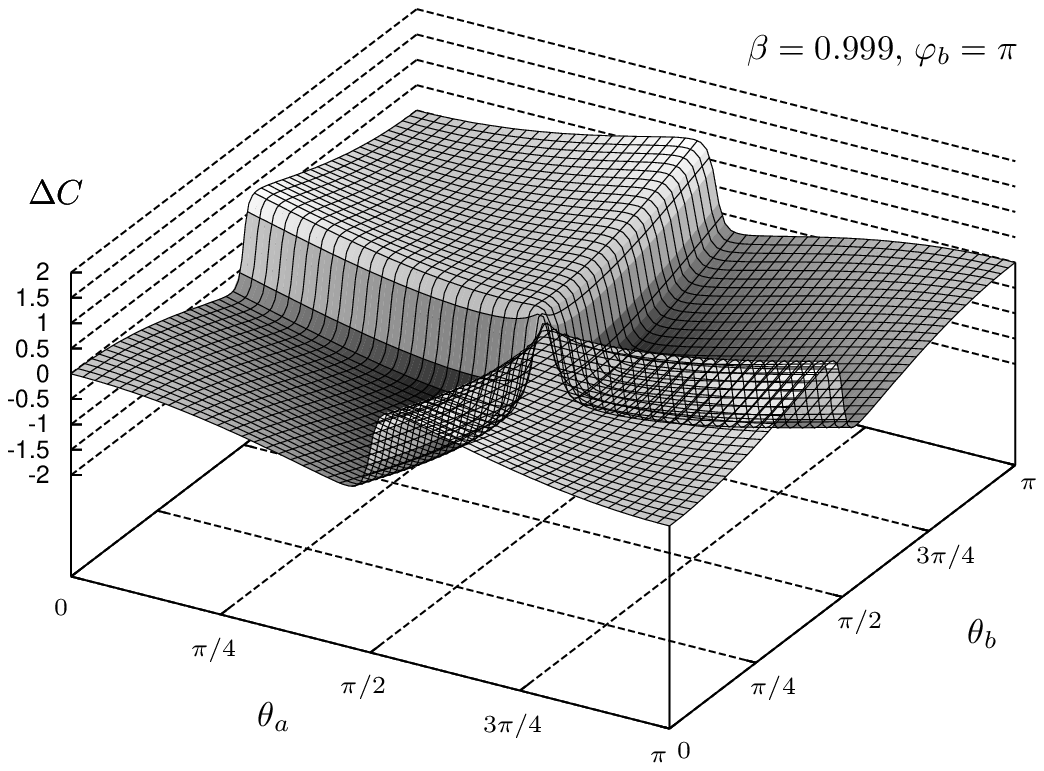}\hfill\rule{0mm}{1mm}
\caption{Function $\Delta C$ (Eq.\ (\ref{cab_correlation_difference}))
  in the parametrization  $\vec{n}=(0,0,1)$,
$\vec{a}=(\sin{\theta_a},0,\cos{\theta_a})$,
$\vec{b}=(\cos{\varphi_b}\sin{\theta_b},\sin{\varphi_b}\sin{\theta_b},
\cos{\theta_b})$.
We have plotted the graph
of the function $\Delta C$ for two arbitrarily chosen values of
$\varphi_b$ and for $\beta=0.999$, which is the value for
electron--positron pair created in the $\pi^0$ decay. }
\label{rys}
\end{figure}

\subsection{Vector state}

General vector state is given by Eq.\ (\ref{cab:state_vector_chosen}).
So in this case from (\ref{cab:correlation_general}) we get
 \begin{multline}
 C_{\ind{vec}}^{\reg{AB}}(\vec{a},\vec{b}) = \\
 \Bigg\{\int \frac{d^3\vec{k}}{2k^0} \frac{d^3\vec{p}}{2p^0}
 \chi_{\reg{A}}(p) \chi_{\reg{B}}(k) \left(\delta^4(q-k-p)\right)^2 
 \varphi_\mu(k,p) \varphi_{\nu}^{*}(k,p) %\times\\
 \text{Tr}\left[\vec{b}{\cdot}\boldsymbol{\sigma} \Big(\vv^{T}(k) \mathcal{C}
   \gamma^\mu \vv(p) \Big) \vec{a}{\cdot}\boldsymbol{\sigma}^T 
  \Big(\vv^{T}(k) \mathcal{C} \gamma^\nu \vv(p) \Big)^\dag
\right]\Bigg\} \\
 \times \Bigg\{\int \frac{d^3\vec{k}}{2k^0} \frac{d^3\vec{p}}{2p^0} 
 \chi_{\reg{A}}(p) \chi_{\reg{B}}(k) \left(\delta^4(q-k-p)\right)^2 %\times \\
 \varphi_\mu(k,p) \varphi_{\nu}^{*}(k,p)\,
 \text{Tr}\left[\Big(\vv^{T}(k) \mathcal{C} \gamma^\mu \vv(p) \Big)  
  \Big(\vv^{T}(k) \mathcal{C} \gamma^\nu \vv(p) \Big)^\dag
 \right]\Bigg\}^{-1} ,
 \label{cab:correl_function_vector_general}
 \end{multline}
where, as before, $\chi_{\reg{A}}(p)$ i $\chi_{\reg{B}}(k)$ are characteristic
functions of the 
regions $\reg{A}$ and $\reg{B}$ in momentum space. 
If the observers frame coincides with 
the center of mass reference frame (in which
$p=k^\pi=(k^0,-\vec{k})$) then by means of the
transversality condition (\ref{cab:transversality_cond}), we have
 \begin{multline}
 \varphi_\mu(k,k^\pi) \varphi_{\nu}^{*}(k,k^\pi) 
 \text{Tr}\left[\vec{b}{\cdot}\boldsymbol{\sigma} \Big(\vv^{T}(k) \mathcal{C}
   \gamma^\mu \vv(k^\pi) \Big) \vec{a}{\cdot}\boldsymbol{\sigma}^T 
  \Big(\vv^{T}(k) \mathcal{C} \gamma^\nu \vv(k^\pi) \Big)^\dag 
 \right] = \\
 = \frac{2}{m^2} \Bigg\{ (\vec{a}{\cdot}\vec{b}) \left[ {k^0}^2
   |\boldsymbol{\varphi}|^2 - |\vec{k}{\cdot}\boldsymbol{\varphi}|^2\right] -
 {k^0}^2 \left[ (\vec{a}{\cdot}\boldsymbol{\varphi}^*) 
 (\vec{b}{\cdot}\boldsymbol{\varphi}) +
 (\vec{a}{\cdot}\boldsymbol{\varphi})
 (\vec{b}{\cdot}\boldsymbol{\varphi}^*) \right] 
 - \frac{2(\vec{a}{\cdot}\vec{k}) (\vec{b}{\cdot}\vec{k}) 
 |\vec{k}{\cdot}\boldsymbol{\varphi}|^2}{(m+k^0)^2} \\ + 
 \frac{k^0}{m+k^0}\Big[ (\vec{a}{\cdot}\vec{k}) 
 \left( (\vec{b}{\cdot}\boldsymbol{\varphi})
 (\boldsymbol{\varphi}^*{\cdot}\vec{k}) +
 (\vec{b}{\cdot}\boldsymbol{\varphi}^*)
 (\boldsymbol{\varphi}{\cdot}\vec{k})\right) + 
 (\vec{b}{\cdot}\vec{k}) \left( (\vec{a}{\cdot}\boldsymbol{\varphi})
 (\boldsymbol{\varphi}^*{\cdot}\vec{k}) +
 (\vec{a}{\cdot}\boldsymbol{\varphi}^*)
 (\boldsymbol{\varphi}{\cdot}\vec{k})\right) \Big] \Bigg\}. 
 \end{multline}
 \begin{equation}
 \varphi_\mu(k,k^\pi) \varphi_{\nu}^{*}(k,k^\pi) 
 \text{Tr}\left[\Big(\vv^{T}(k) \mathcal{C} \gamma^\mu \vv(k^\pi) \Big)
 \Big(\vv^{T}(k) \mathcal{C} \gamma^\nu \vv(k^\pi) \Big)^\dag 
 \right] 
 = \frac{2}{m^2} \Big[ {k^0}^2 |\boldsymbol{\varphi}|^2 - 
 |\vec{k}{\cdot}\boldsymbol{\varphi}|^2 \Big].
 \end{equation}
Therefore, if we assume that momenta of the particles in the state
$\ket{\varphi}^{\ind{vec}}_{q}$ are sharp, the
correlation function in the center of mass frame
are given by (antiparticle has momentum $k$,
particle $k^\pi$)
 \begin{multline}
 C_{{\ind{vec}}}^{\ind{CMF}}(\vec{a},\vec{b}) = 
 \vec{a}{\cdot}\vec{b} - \frac{{k^0}^2}{{k^0}^2
   |\boldsymbol{\varphi}|^2 - |\vec{k}{\cdot}\boldsymbol{\varphi}|^2}
 \left[(\vec{a}{\cdot}\boldsymbol{\varphi}^*) 
 (\vec{b}{\cdot}\boldsymbol{\varphi}) + 
 (\vec{a}{\cdot}\boldsymbol{\varphi})
 (\vec{b}{\cdot}\boldsymbol{\varphi}^*) \right] 
 - \frac{2(\vec{a}{\cdot}\vec{k})(\vec{b}{\cdot}\vec{k})
 |\vec{k}{\cdot}\boldsymbol{\varphi}|^2}%
 {(m+k^0)^2 \big({k^0}^2|\boldsymbol{\varphi}|^2 - 
 |\vec{k}{\cdot}\boldsymbol{\varphi}|^2
   \big)} \\
 + \frac{k^0
 \left[ (\vec{a}{\cdot}\vec{k}) \left( (\vec{b}{\cdot}\boldsymbol{\varphi})
 (\vec{k}{\cdot}\boldsymbol{\varphi}^*) +
 (\vec{b}{\cdot}\boldsymbol{\varphi}^*)
 (\vec{k}{\cdot}\boldsymbol{\varphi})\right) + 
 (\vec{b}{\cdot}\vec{k}) \left( (\vec{a}{\cdot}\boldsymbol{\varphi})
 (\vec{k}{\cdot}\boldsymbol{\varphi}^*) +
 (\vec{a}{\cdot}\boldsymbol{\varphi}^*)
 (\vec{k}{\cdot}\boldsymbol{\varphi})\right)\right]}%
 {(m+k^0) \big({k^0}^2|\boldsymbol{\varphi}|^2 - 
 |\vec{k}{\cdot}\boldsymbol{\varphi}|^2
   \big)}. \label{cab:triplet_corr_CMS}
 \end{multline}
Notice that in this case, in opposition to the one considered previously, in
the center of mass frame the correlation function depends explicitly on
the momentum. Moreover, if $\vec{k}\perp\boldsymbol{\varphi}$ from
(\ref{cab:triplet_corr_CMS}) we get the same result as for 
nonrelativistic triplet state (\ref{cab:triplet_nonrel_corr}).
Also in the nonrelativistic limit (${k^0}^2 \gg |\vec{k}|^2$) from
(\ref{cab:triplet_corr_CMS}) we get the correlation function for
nonrelativistic triplet state (\ref{cab:triplet_nonrel_corr}).

When we consider electron--positron pairs produced in the $Z^0$ decay
two remarks are in order. First, in some experiments it is possible to
produce polarized $Z^0$ (SLAC, Stanford\cite{cab_SLAC}), so in such
experiments the vector 
$\boldsymbol{\varphi}$ is determined. Second, electron and positrons
produced in the channel $Z^0\to e^+e^-$ are highly
ultrarelativistic. In the limit $\beta\to1$ the formula
(\ref{cab:triplet_corr_CMS}) takes the form
 \begin{multline}
 \left.C_{{\ind{vec}}}^{\ind{CMF}}(\vec{a},\vec{b})\right|_{\beta\to1} = 
 \vec{a}{\cdot}\vec{b} - \frac{1}{
   |\boldsymbol{\varphi}|^2 - |\vec{n}{\cdot}\boldsymbol{\varphi}|^2}
 \Big\{(\vec{a}{\cdot}\boldsymbol{\varphi}^*) 
 (\vec{b}{\cdot}\boldsymbol{\varphi}) + 
 (\vec{a}{\cdot}\boldsymbol{\varphi})(\vec{b}{\cdot}\boldsymbol{\varphi}^*)  
 + 2(\vec{a}{\cdot}\vec{n})(\vec{b}{\cdot}\vec{n})
 |\vec{n}{\cdot}\boldsymbol{\varphi}|^2 \\
 - (\vec{a}{\cdot}\vec{n}) 
 \big[ (\vec{b}{\cdot}\boldsymbol{\varphi}) 
 (\vec{n}{\cdot}\boldsymbol{\varphi}^*) +
 (\vec{b}{\cdot}\boldsymbol{\varphi}^*)
 (\vec{n}{\cdot}\boldsymbol{\varphi})\big]  
 - (\vec{b}{\cdot}\vec{n}) 
 \big[ (\vec{a}{\cdot}\boldsymbol{\varphi}) 
 (\vec{n}{\cdot}\boldsymbol{\varphi}^*) +
 (\vec{a}{\cdot}\boldsymbol{\varphi}^*)
 (\vec{n}{\cdot}\boldsymbol{\varphi})\big]
 \Big\}, \label{cab:triplet_corr_CMS_ultrarelativistic}
 \end{multline}
where $\vec{n}=\vec{k}/|\vec{k}|$. Eq.\
(\ref{cab:triplet_corr_CMS_ultrarelativistic}) takes very simple form
in the configuration $\vec{a}\perp\boldsymbol{\varphi}$,
$\vec{b}\perp\boldsymbol{\varphi}$:
 \begin{equation}
 \left.C_{{\ind{vec}}}^{\ind{CMF}}(\vec{a}\perp\boldsymbol{\varphi},
 \vec{b}\perp\boldsymbol{\varphi})\right|_{\beta\to1} = 
 \vec{a}{\cdot}\vec{b} - 
 \frac{2(\vec{a}{\cdot}\vec{n})(\vec{b}{\cdot}\vec{n})
 |\vec{n}{\cdot}\boldsymbol{\varphi}|^2}%
{|\boldsymbol{\varphi}|^2 -|\vec{n}{\cdot}\boldsymbol{\varphi}|^2},
 \label{cab:triplet_corr_CMS_ultrarelativistic_specific}
 \end{equation}
It is interesting to notice, that the function
 \begin{equation}
 \Delta C_{{\ind{vec}}} =
 \left.C_{{\ind{vec}}}^{\ind{CMF}}(\vec{a}\perp\boldsymbol{\varphi},
 \vec{b}\perp\boldsymbol{\varphi})\right|_{\beta\to1} - 
 C_{{\ind{nonrelativ}}} =  - 
 \frac{2(\vec{a}{\cdot}\vec{n})(\vec{b}{\cdot}\vec{n})
 |\vec{n}{\cdot}\boldsymbol{\varphi}|^2}%
{|\boldsymbol{\varphi}|^2 -|\vec{n}{\cdot}\boldsymbol{\varphi}|^2},
 \label{cab_correlation_difference_vector}
 \end{equation}   
where $C_{{\ind{nonrelativ}}}$ denotes correlation function for the
nonrelativistic triplet state (\ref{cab:triplet_nonrel_corr}), can
take arbitrary value from the interval $[-2,2]$. We shown this 
function in the Fig.\ \ref{rys2}.
\end{widetext}

\begin{figure}
\rule{0mm}{1mm}\hfill
\includegraphics[width=.95\columnwidth]{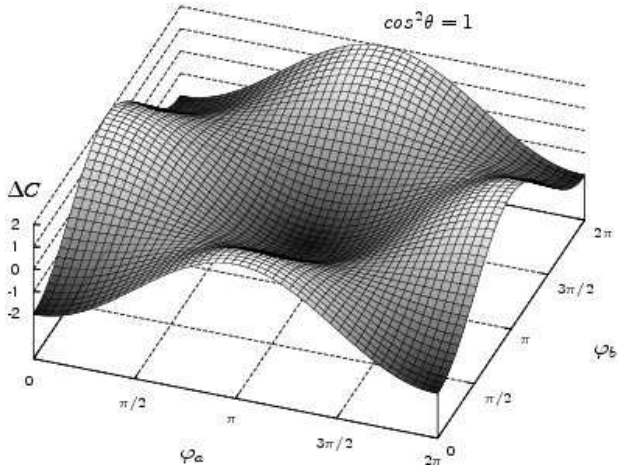}\hfill\rule{0mm}{1mm}
\caption{Function $\Delta C$ (Eq.\ (\ref{cab_correlation_difference_vector}))
  in the parametrization  $\boldsymbol{\varphi}=(0,0,1)$, 
$\vec{n}=(\sin\theta,0,\cos\theta)$, 
$\vec{a}=(\cos{\varphi_a},\sin{\varphi_a},0)$,
$\vec{b}= (\cos{\varphi_b},\sin{\varphi_b},0)$. 
In this parametrization $\Delta C= -2 \cos{\varphi_a} \cos{\varphi_b}
\cos^2\theta$. We have plotted the graph
for the value $\theta$ such that
$\cos^2\theta=1$. We see that $\Delta C$ is of the same order of
magnitude as the correlation function. }
\label{rys2}
\end{figure}

\section{Conclusions}

We have discussed correlations of two relativistic massive particles
in the EPR--type experiments in the context of quantum field theory
formalism. 
Choosing the most appropriate
spin operator (\ref{spin}), we have calculated correlation function
for the particle--anti--particle pair in the pseudoscalar and vector
states.  We have found general formulas and discussed them in some
details for the sharp momentum states in the specific configurations.
Relativistic correlations in the vector state have been never
discussed before. It should also be noted that all these functions
posses a proper nonrelativistic limit. 

It is interesting that for the pseudoscalar (singlet) state the
correlation function in the center of mass frame is the same as in the
nonrelativistic case. Therefore, in opposition to Czachor's results
\cite{Czachor1997_1}, the degree of violation of Bell inequality for
the pseudoscalar state in CMF is independent of the particle
momentum. The correlation function for the vector state depends on the
configuration and for same configurations gives the same result as for
the nonrelativistic triplet state.

It seems that EPR--type experiment with relativistic elementary
particles are feasible with the present technology
\cite{cab_Ciborowski2006}.

\begin{acknowledgments}
This paper has been partially supported by the Polish Ministry of
Scientific Research and Information Technology under Grant No.\ 
PBZ-MIN-008/P03/2003 and partially by the University of Lodz grant.
\end{acknowledgments}

\appendix
\section{Dirac matrices}
\label{app:Dirac_matrices}
In this paper we use the following conventions.
Dirac matrices fulfill the relation
$\gamma^\mu\gamma^\nu+\gamma^\nu\gamma^\mu=2g^{\mu\nu}$ where
the Minkowski metric tensor $g^{\mu\nu}=\text{diag}(1,-1,-1,-1)$;
moreover we adopt the convention $\varepsilon^{0123}=1$.  We use the
following explicit representation of gamma matrices:
 \protect\begin{equation}
 \gamma^0=\left(\protect\begin{array}{cc}
 0 & I \\ I & 0
 \protect\end{array}\right),\quad 
 {\boldsymbol{\gamma}}=\left(\protect\begin{array}{cc}
 0 & -{\boldsymbol{\sigma}} \\ {\boldsymbol{\sigma}} & 0
 \protect\end{array}\right),\quad \gamma^5= 
 \left(\protect\begin{array}{cc} 
 I & 0 \\ 0 & -I
 \protect\end{array}\right),
 \label{gamma_explicit}
 \protect\end{equation} 
where $\boldsymbol{\sigma}=(\sigma_1,\sigma_2,\sigma_3)$ and
$\sigma_i$ are standard Pauli matrices. The charge conjugation matrix
has the form
 \begin{equation}
 \mathcal{C} = -i \gamma^2 \gamma^0 = i
 \left(\begin{smallmatrix} \sigma_2 & 0 \\ 0 & -\sigma_2
   \end{smallmatrix}\right),
 \label{cab:charge_conjug}
 \end{equation}
so
 \begin{equation}
 \mathcal{C} \gamma^{T\mu} \mathcal{C}^{-1} = - \gamma^\mu.
 \label{cab:charge_conjug_condition}
 \end{equation}

\section{Bispinor representation}
\label{cab:app_bispinor_rep}

By the bispinor representation of the Lorentz group we mean the
representation 
$D^{(\frac{1}{2},0)}\oplus D^{(0,\frac{1}{2})}$. Explicitly, if $A\in
SL(2,{\mathbb{C}})$ and $\Lambda(A)$ is an image of $A$ in the
canonical homomorphism
of the $SL(2,{\mathbb{C}})$ group onto the
Lorentz group, we take the chiral form of $D^{(\frac{1}{2},0)}\oplus
D^{(0,\frac{1}{2})}$, namely 
 \begin{equation}
 \rep{D}(\Lambda(A))= \begin{pmatrix}
 A & 0 \\ 0 & (A^\dag)^{-1}
 \end{pmatrix}.
 \label{representation}
 \end{equation}
The canonical homomorphism between the group
$SL(2,{\mathbb{C}})$ (universal covering of the proper ortochronous
Lorentz group $L_{+}^{\uparrow}$) and the Lorentz group
$L_{+}^{\uparrow}\sim SO(1,3)_0$ \cite{cab_BR1977}  
is defined as follows:
With every four-vector $k^\mu$ we associate a two-dimensional
hermitian matrix $k^\mu \sigma_\mu$,
where $\sigma_i$, $i=1,2,3$, are the standard Pauli matrices
and $\sigma_0=I$. In the space of two-dimensional hermitian matrices
the Lorentz group action is given by 
${{\sf k}^\prime}^\mu \sigma_\mu = A ({\textsf k}^\mu \sigma_\mu)
A^\dag$, where $A$ denotes the element of the  
$SL(2,{\mathbb{C}})$ group corresponding to the Lorentz transformation 
$\Lambda(A)$ which converts the four-vector $k$ to $k^\prime$ (i.e.,
${k^\prime}^\mu=\Lambda_{\phantom{\mu}\nu}^{\mu}k^\nu$).

\section{Useful formulas} 

The following relations hold:
\begin{align}
  & (k\gamma)\, \vv(k) =  m \vv(k), \\
  & \bar{\vv}(k) \vv(k) = I, \\
  & \gamma^0 \vv(k) = \vv(k^\pi), \\
  & \bar{\vv}(k) \gamma^\mu \vv(k) = \frac{k^\mu}{m} I,\\
  & \vv(k) \bar{\vv}(k) = \tfrac{1}{2m}(k\gamma+mI),
 \label{cab:amplitudes_connection}
 \end{align}
and
 \begin{gather}
  \vv(\Lambda k) = \textsf{D}(\Lambda) \vv(k)
  \mathcal{D}^T(R(\Lambda,k)),
  \label{cab:Weinberg_cond}\\
  \vv(\Lambda k) \sigma_2 = \textsf{D}(\Lambda)\, \vv(k)\sigma_2\,
  \mathcal{D}^\dag(R(\Lambda,k)).  
 \end{gather} 
Notice that when operator $\Hat{A}$ acts on standard basis vectors in
the following way
 \begin{equation}
 \Hat{A} \ket{k,\sigma} = A_{\sigma\sigma^\prime}
 \ket{k,\sigma^\prime}, 
 \end{equation}
then its action on the covariant basis (\ref{cab:covariant_vectors})
is of the form
 \begin{equation}
 \Hat{A} \kket{\alpha,k} = \left( \vv(k) A \bar{\vv}(k)
 \right)_{\alpha\alpha^\prime} \kket{\alpha^\prime,k}.
 \end{equation}
Analogously for two--particle states the relation
 \begin{equation}
 (\Hat{A}\otimes\Hat{B}) \ket{(k,\sigma)_b,(p,\lambda)_a} = 
 A_{\sigma\sigma^\prime} B_{\lambda\lambda^\prime}
 \ket{(k,\sigma^\prime)_b,(p,\lambda^\prime)_a}
 \end{equation}
implies
 \begin{multline}
 (\Hat{A}\otimes\Hat{B}) \kket{(\alpha,k),(\beta,p)}
 = \left( \vv(k) \Hat{A} \bar{\vv}(k)
 \right)_{\alpha\alpha^\prime}  \\
 \left( \vv(p) \Hat{B} \bar{\vv}(p) \right)_{\beta\beta^\prime} 
 \kket{(\alpha^\prime,k),(\beta^\prime,p)}.
 \end{multline}

\section{Correlations in nonrelativistic triplet state}

We recall in this appendix the correlation function in nonrelativistic
triplet 
state (see e.g.\ \cite{cab_CRSW2003_1}). Let $\ket{\sigma}$,
$\sigma=\pm\frac{1}{2}$, denotes eigenvector of the 
spin component along $z$ axis corresponding to the value of spin
z--component equal to $\sigma$. The general triplet state has the form 
 \begin{equation}
 \ket{\varphi} = \sum\limits_{\sigma,\lambda=\pm1/2} \varphi_{\sigma
   \lambda} \ket{\sigma}\otimes\ket{\lambda},
 \label{cab:triplet_nonrel}
 \end{equation}
with the symmetry condition
 \begin{equation}
 \varphi_{\sigma \lambda} = \varphi_{\lambda \sigma}. 
 \end{equation}
It is convenient to parametrize matrix $\varphi=[\varphi_{\sigma
  \lambda}]$ in the following way:
 \begin{equation}
 \varphi = \frac{i}{\sqrt{2}} 
 ({\boldsymbol{\varphi}}{\cdot}{\boldsymbol{\sigma}}) \sigma_2.
 \end{equation}
%where normalization of the state (\ref{cab:triplet_nonrel}) yields
%$|{\boldsymbol{\varphi}}|^2=1$. 
So finally the normalized correlation
function in the triplet state is given by
 \begin{eqnarray}
 C(\vec{a},\vec{b}) & = & \bra{\varphi}
 \vec{a}{\cdot}{\boldsymbol{\sigma}}\otimes\vec{b}{\cdot}{\boldsymbol{\sigma}}
 \ket{\varphi} \\ 
  & = &  \vec{a}{\cdot}\vec{b} - \frac{1}{|{\boldsymbol{\varphi}}|^2}
 [(\vec{a}{\cdot}{\boldsymbol{\varphi}})
 (\vec{b}{\cdot}{\boldsymbol{\varphi}}^*) +
 (\vec{b}{\cdot}{\boldsymbol{\varphi}})
 (\vec{a}{\cdot}{\boldsymbol{\varphi}}^*) ] .
 \label{cab:triplet_nonrel_corr}
 \end{eqnarray}

%\bibliography{quantum,rel_entanglement_for_triplet}

\begin{thebibliography}{51}
\expandafter\ifx\csname natexlab\endcsname\relax\def\natexlab#1{#1}\fi
\expandafter\ifx\csname bibnamefont\endcsname\relax
  \def\bibnamefont#1{#1}\fi
\expandafter\ifx\csname bibfnamefont\endcsname\relax
  \def\bibfnamefont#1{#1}\fi
\expandafter\ifx\csname citenamefont\endcsname\relax
  \def\citenamefont#1{#1}\fi
\expandafter\ifx\csname url\endcsname\relax
  \def\url#1{\texttt{#1}}\fi
\expandafter\ifx\csname urlprefix\endcsname\relax\def\urlprefix{URL }\fi
\providecommand{\bibinfo}[2]{#2}
\providecommand{\eprint}[2][]{\url{#2}}

\bibitem[{\citenamefont{Einstein et~al.}(1935)\citenamefont{Einstein, Podolsky,
  and Rosen}}]{cab_EPR1935}
\bibinfo{author}{\bibfnamefont{A.}~\bibnamefont{Einstein}},
  \bibinfo{author}{\bibfnamefont{B.}~\bibnamefont{Podolsky}}, \bibnamefont{and}
  \bibinfo{author}{\bibfnamefont{N.}~\bibnamefont{Rosen}},
  \bibinfo{journal}{Phys.\ Rev.} \textbf{\bibinfo{volume}{47}},
  \bibinfo{pages}{777} (\bibinfo{year}{1935}).

\bibitem[{\citenamefont{Rove et~al.}(2001)\citenamefont{Rove, Kielpinski,
  Meyer, Sackett, Itano, Monroe, and Wineland}}]{cab_RKMSIMW2001}
\bibinfo{author}{\bibfnamefont{M.~A.} \bibnamefont{Rove}},
  \bibinfo{author}{\bibfnamefont{D.}~\bibnamefont{Kielpinski}},
  \bibinfo{author}{\bibfnamefont{V.}~\bibnamefont{Meyer}},
  \bibinfo{author}{\bibfnamefont{C.~A.} \bibnamefont{Sackett}},
  \bibinfo{author}{\bibfnamefont{W.~M.} \bibnamefont{Itano}},
  \bibinfo{author}{\bibfnamefont{C.}~\bibnamefont{Monroe}}, \bibnamefont{and}
  \bibinfo{author}{\bibfnamefont{D.~J.} \bibnamefont{Wineland}},
  \bibinfo{journal}{Nature} \textbf{\bibinfo{volume}{409}},
  \bibinfo{pages}{781} (\bibinfo{year}{2001}).

\bibitem[{\citenamefont{Weihs et~al.}(1998)\citenamefont{Weihs, Jennewein,
  Simon, Weinfurter, and Zeilinger}}]{cab_Weihs1998}
\bibinfo{author}{\bibfnamefont{G.}~\bibnamefont{Weihs}},
  \bibinfo{author}{\bibfnamefont{T.}~\bibnamefont{Jennewein}},
  \bibinfo{author}{\bibfnamefont{C.}~\bibnamefont{Simon}},
  \bibinfo{author}{\bibfnamefont{H.}~\bibnamefont{Weinfurter}},
  \bibnamefont{and}
  \bibinfo{author}{\bibfnamefont{A.}~\bibnamefont{Zeilinger}},
  \bibinfo{journal}{Phys. Rev. Lett.} \textbf{\bibinfo{volume}{81}},
  \bibinfo{pages}{5039} (\bibinfo{year}{1998}).

\bibitem[{\citenamefont{Czachor}(1997{\natexlab{a}})}]{Czachor1997_1}
\bibinfo{author}{\bibfnamefont{M.}~\bibnamefont{Czachor}},
  \bibinfo{journal}{Phys. Rev. A} \textbf{\bibinfo{volume}{55}},
  \bibinfo{pages}{72} (\bibinfo{year}{1997}{\natexlab{a}}).

\bibitem[{\citenamefont{Czachor}(1997{\natexlab{b}})}]{Czachor1997_2}
\bibinfo{author}{\bibfnamefont{M.}~\bibnamefont{Czachor}}, in
  \emph{\bibinfo{booktitle}{Photonic Quantum Computing}}, edited by
  \bibinfo{editor}{\bibfnamefont{S.~P.} \bibnamefont{Holating}}
  \bibnamefont{and} \bibinfo{editor}{\bibfnamefont{A.~R.} \bibnamefont{Pirich}}
  (\bibinfo{publisher}{SPIE - The International Society for Optical
  Engineering}, \bibinfo{address}{Bellingham, WA},
  \bibinfo{year}{1997}{\natexlab{b}}), vol. \bibinfo{volume}{3076} of
  \emph{\bibinfo{series}{Proceedings of SPIE}}, pp. \bibinfo{pages}{141--145}.

\bibitem[{\citenamefont{Ahn et~al.}(2002)\citenamefont{Ahn, Lee, and
  Hwang}}]{ALH2002}
\bibinfo{author}{\bibfnamefont{D.}~\bibnamefont{Ahn}},
  \bibinfo{author}{\bibfnamefont{H.~J.} \bibnamefont{Lee}}, \bibnamefont{and}
  \bibinfo{author}{\bibfnamefont{S.~W.} \bibnamefont{Hwang}}
  (\bibinfo{year}{2002}), \eprint{arXiv: quant-ph/0207018}.

\bibitem[{\citenamefont{Ahn et~al.}(2003{\natexlab{a}})\citenamefont{Ahn, Lee,
  Hwang, and Kim}}]{ALHK2003}
\bibinfo{author}{\bibfnamefont{D.}~\bibnamefont{Ahn}},
  \bibinfo{author}{\bibfnamefont{H.~J.} \bibnamefont{Lee}},
  \bibinfo{author}{\bibfnamefont{S.~W.} \bibnamefont{Hwang}}, \bibnamefont{and}
  \bibinfo{author}{\bibfnamefont{M.~S.} \bibnamefont{Kim}}
  (\bibinfo{year}{2003}{\natexlab{a}}), \eprint{arXiv: quant-ph/0304119}.

\bibitem[{\citenamefont{Ahn et~al.}(2003{\natexlab{b}})\citenamefont{Ahn, Lee,
  Moon, and Hwang}}]{ALMH2003}
\bibinfo{author}{\bibfnamefont{D.}~\bibnamefont{Ahn}},
  \bibinfo{author}{\bibfnamefont{H.~J.} \bibnamefont{Lee}},
  \bibinfo{author}{\bibfnamefont{Y.~H.} \bibnamefont{Moon}}, \bibnamefont{and}
  \bibinfo{author}{\bibfnamefont{S.~W.} \bibnamefont{Hwang}},
  \bibinfo{journal}{Phys. Rev. A} \textbf{\bibinfo{volume}{67}},
  \bibinfo{pages}{012103} (\bibinfo{year}{2003}{\natexlab{b}}).

\bibitem[{\citenamefont{Alsing and Milburn}(2002)}]{AM2002}
\bibinfo{author}{\bibfnamefont{P.~M.} \bibnamefont{Alsing}} \bibnamefont{and}
  \bibinfo{author}{\bibfnamefont{G.~J.} \bibnamefont{Milburn}},
  \bibinfo{journal}{Q. Inf. and Comput.} \textbf{\bibinfo{volume}{2}},
  \bibinfo{pages}{487} (\bibinfo{year}{2002}).

\bibitem[{\citenamefont{Bartlett and Terno}(2005)}]{BT2005}
\bibinfo{author}{\bibfnamefont{S.~D.} \bibnamefont{Bartlett}} \bibnamefont{and}
  \bibinfo{author}{\bibfnamefont{D.~R.} \bibnamefont{Terno}},
  \bibinfo{journal}{Phys. Rev. A} \textbf{\bibinfo{volume}{71}},
  \bibinfo{pages}{012302} (\bibinfo{year}{2005}).

\bibitem[{\citenamefont{Caban and Rembieli\'nski}(2003)}]{CR2003_Wigner}
\bibinfo{author}{\bibfnamefont{P.}~\bibnamefont{Caban}} \bibnamefont{and}
  \bibinfo{author}{\bibfnamefont{J.}~\bibnamefont{Rembieli\'nski}},
  \bibinfo{journal}{Phys. Rev. A} \textbf{\bibinfo{volume}{68}},
  \bibinfo{pages}{042107} (\bibinfo{year}{2003}).

\bibitem[{\citenamefont{Caban and Rembieli\'nski}(2005)}]{CR2005}
\bibinfo{author}{\bibfnamefont{P.}~\bibnamefont{Caban}} \bibnamefont{and}
  \bibinfo{author}{\bibfnamefont{J.}~\bibnamefont{Rembieli\'nski}},
  \bibinfo{journal}{Phys. Rev. A} \textbf{\bibinfo{volume}{72}},
  \bibinfo{pages}{012103} (\bibinfo{year}{2005}).

\bibitem[{\citenamefont{Czachor and Wilczewski}(2003)}]{CW2003}
\bibinfo{author}{\bibfnamefont{M.}~\bibnamefont{Czachor}} \bibnamefont{and}
  \bibinfo{author}{\bibfnamefont{M.}~\bibnamefont{Wilczewski}},
  \bibinfo{journal}{Phys. Rev. A} \textbf{\bibinfo{volume}{68}},
  \bibinfo{pages}{010302(R)} (\bibinfo{year}{2003}).

\bibitem[{\citenamefont{Czachor}(2005)}]{Czachor2005}
\bibinfo{author}{\bibfnamefont{M.}~\bibnamefont{Czachor}},
  \bibinfo{journal}{Phys. Rev. Lett.} \textbf{\bibinfo{volume}{94}},
  \bibinfo{pages}{078901} (\bibinfo{year}{2005}).

\bibitem[{\citenamefont{Gingrich and Adami}(2002)}]{GA2002}
\bibinfo{author}{\bibfnamefont{R.~M.} \bibnamefont{Gingrich}} \bibnamefont{and}
  \bibinfo{author}{\bibfnamefont{C.}~\bibnamefont{Adami}},
  \bibinfo{journal}{Phys. Rev. Lett.} \textbf{\bibinfo{volume}{89}},
  \bibinfo{pages}{270402} (\bibinfo{year}{2002}).

\bibitem[{\citenamefont{Gingrich et~al.}(2003)\citenamefont{Gingrich, Bergou,
  and Adami}}]{GBA2003}
\bibinfo{author}{\bibfnamefont{R.~M.} \bibnamefont{Gingrich}},
  \bibinfo{author}{\bibfnamefont{A.~J.} \bibnamefont{Bergou}},
  \bibnamefont{and} \bibinfo{author}{\bibfnamefont{C.}~\bibnamefont{Adami}},
  \bibinfo{journal}{Phys. Rev. A} \textbf{\bibinfo{volume}{68}},
  \bibinfo{pages}{042102} (\bibinfo{year}{2003}).

\bibitem[{\citenamefont{Gonera et~al.}(2004)\citenamefont{Gonera, Kosi\'nski,
  and Ma\'slanka}}]{GKM2004}
\bibinfo{author}{\bibfnamefont{C.}~\bibnamefont{Gonera}},
  \bibinfo{author}{\bibfnamefont{P.}~\bibnamefont{Kosi\'nski}},
  \bibnamefont{and}
  \bibinfo{author}{\bibfnamefont{P.}~\bibnamefont{Ma\'slanka}},
  \bibinfo{journal}{Phys. Rev. A} \textbf{\bibinfo{volume}{70}},
  \bibinfo{pages}{034102} (\bibinfo{year}{2004}).

\bibitem[{\citenamefont{Harshman}(2005)}]{Harshman2005}
\bibinfo{author}{\bibfnamefont{N.~L.} \bibnamefont{Harshman}},
  \bibinfo{journal}{Phys. Rev. A} \textbf{\bibinfo{volume}{71}},
  \bibinfo{pages}{022312} (\bibinfo{year}{2005}).

\bibitem[{\citenamefont{Jordan et~al.}(2005)\citenamefont{Jordan, Shaji, and
  Sudarshan}}]{JSS2005}
\bibinfo{author}{\bibfnamefont{T.~F.} \bibnamefont{Jordan}},
  \bibinfo{author}{\bibfnamefont{A.}~\bibnamefont{Shaji}}, \bibnamefont{and}
  \bibinfo{author}{\bibfnamefont{E.~C.~G.} \bibnamefont{Sudarshan}},
  \bibinfo{journal}{Phys. Rev. A} \textbf{\bibinfo{volume}{73}},
  \bibinfo{pages}{032104} (\bibinfo{year}{2005}).

\bibitem[{\citenamefont{Kosi\'nski and Ma\'slanka}(2003)}]{KM2003}
\bibinfo{author}{\bibfnamefont{P.}~\bibnamefont{Kosi\'nski}} \bibnamefont{and}
  \bibinfo{author}{\bibfnamefont{P.}~\bibnamefont{Ma\'slanka}}
  (\bibinfo{year}{2003}), \eprint{arXiv: quant-ph/0310145}.

\bibitem[{\citenamefont{Lamata et~al.}(2005)\citenamefont{Lamata,
  Martin-Delgado, and Solano}}]{LMS2005}
\bibinfo{author}{\bibfnamefont{L.}~\bibnamefont{Lamata}},
  \bibinfo{author}{\bibfnamefont{M.~A.} \bibnamefont{Martin-Delgado}},
  \bibnamefont{and} \bibinfo{author}{\bibfnamefont{E.}~\bibnamefont{Solano}}
  (\bibinfo{year}{2005}), \eprint{arXiv: quant-ph/051208}.

\bibitem[{\citenamefont{Lee and Chang-Young}(2004)}]{LY2004}
\bibinfo{author}{\bibfnamefont{D.}~\bibnamefont{Lee}} \bibnamefont{and}
  \bibinfo{author}{\bibfnamefont{E.}~\bibnamefont{Chang-Young}},
  \bibinfo{journal}{New J. Phys.} \textbf{\bibinfo{volume}{6}},
  \bibinfo{pages}{67} (\bibinfo{year}{2004}).

\bibitem[{\citenamefont{Li and Du}(2003)}]{LD2003}
\bibinfo{author}{\bibfnamefont{H.}~\bibnamefont{Li}} \bibnamefont{and}
  \bibinfo{author}{\bibfnamefont{J.}~\bibnamefont{Du}}, \bibinfo{journal}{Phys.
  Rev. A} \textbf{\bibinfo{volume}{68}}, \bibinfo{pages}{022108}
  (\bibinfo{year}{2003}).

\bibitem[{\citenamefont{Li and Du}(2004)}]{LD2004}
\bibinfo{author}{\bibfnamefont{H.}~\bibnamefont{Li}} \bibnamefont{and}
  \bibinfo{author}{\bibfnamefont{J.}~\bibnamefont{Du}}, \bibinfo{journal}{Phys.
  Rev. A} \textbf{\bibinfo{volume}{70}}, \bibinfo{pages}{012111}
  (\bibinfo{year}{2004}).

\bibitem[{\citenamefont{Lindner et~al.}(2003)\citenamefont{Lindner, Peres, and
  Terno}}]{LPT2003}
\bibinfo{author}{\bibfnamefont{N.~H.} \bibnamefont{Lindner}},
  \bibinfo{author}{\bibfnamefont{A.}~\bibnamefont{Peres}}, \bibnamefont{and}
  \bibinfo{author}{\bibfnamefont{D.~R.} \bibnamefont{Terno}},
  \bibinfo{journal}{J. Phys. A: Math. Gen.} \textbf{\bibinfo{volume}{36}},
  \bibinfo{pages}{L449} (\bibinfo{year}{2003}).

\bibitem[{\citenamefont{Moon et~al.}(2003)\citenamefont{Moon, Ahn, and
  Hwang}}]{MAH2003}
\bibinfo{author}{\bibfnamefont{Y.~H.} \bibnamefont{Moon}},
  \bibinfo{author}{\bibfnamefont{D.}~\bibnamefont{Ahn}}, \bibnamefont{and}
  \bibinfo{author}{\bibfnamefont{S.~W.} \bibnamefont{Hwang}}
  (\bibinfo{year}{2003}), \eprint{arXiv: quant-ph/0304116}.

\bibitem[{\citenamefont{Peres et~al.}(2002)\citenamefont{Peres, Scudo, and
  Terno}}]{PST2002}
\bibinfo{author}{\bibfnamefont{A.}~\bibnamefont{Peres}},
  \bibinfo{author}{\bibfnamefont{P.~F.} \bibnamefont{Scudo}}, \bibnamefont{and}
  \bibinfo{author}{\bibfnamefont{D.~R.} \bibnamefont{Terno}},
  \bibinfo{journal}{Phys. Rev. Lett.} \textbf{\bibinfo{volume}{88}},
  \bibinfo{pages}{230402} (\bibinfo{year}{2002}).

\bibitem[{\citenamefont{Peres and Terno}(2003{\natexlab{a}})}]{PT2003_2}
\bibinfo{author}{\bibfnamefont{A.}~\bibnamefont{Peres}} \bibnamefont{and}
  \bibinfo{author}{\bibfnamefont{D.~R.} \bibnamefont{Terno}},
  \bibinfo{journal}{J. Mod. Optics} \textbf{\bibinfo{volume}{50}},
  \bibinfo{pages}{1165} (\bibinfo{year}{2003}{\natexlab{a}}).

\bibitem[{\citenamefont{Peres and Terno}(2003{\natexlab{b}})}]{PT2003_3}
\bibinfo{author}{\bibfnamefont{A.}~\bibnamefont{Peres}} \bibnamefont{and}
  \bibinfo{author}{\bibfnamefont{D.~R.} \bibnamefont{Terno}},
  \bibinfo{journal}{Int. J. Q. Inf.} \textbf{\bibinfo{volume}{1}},
  \bibinfo{pages}{225} (\bibinfo{year}{2003}{\natexlab{b}}).

\bibitem[{\citenamefont{Peres and Terno}(2004)}]{PT2004_1}
\bibinfo{author}{\bibfnamefont{A.}~\bibnamefont{Peres}} \bibnamefont{and}
  \bibinfo{author}{\bibfnamefont{D.~R.} \bibnamefont{Terno}},
  \bibinfo{journal}{Rev. Mod. Phys.} \textbf{\bibinfo{volume}{76}},
  \bibinfo{pages}{93} (\bibinfo{year}{2004}).

\bibitem[{\citenamefont{Peres et~al.}(2005)\citenamefont{Peres, Scudo, and
  Terno}}]{PST2005}
\bibinfo{author}{\bibfnamefont{A.}~\bibnamefont{Peres}},
  \bibinfo{author}{\bibfnamefont{P.~F.} \bibnamefont{Scudo}}, \bibnamefont{and}
  \bibinfo{author}{\bibfnamefont{D.~R.} \bibnamefont{Terno}},
  \bibinfo{journal}{Phys. Rev. Lett.} \textbf{\bibinfo{volume}{94}},
  \bibinfo{pages}{078902} (\bibinfo{year}{2005}).

\bibitem[{\citenamefont{Pachos and Solano}(2003)}]{PS2003}
\bibinfo{author}{\bibfnamefont{J.}~\bibnamefont{Pachos}} \bibnamefont{and}
  \bibinfo{author}{\bibfnamefont{E.}~\bibnamefont{Solano}},
  \bibinfo{journal}{Q. Inf. and Comput.} \textbf{\bibinfo{volume}{3}},
  \bibinfo{pages}{115} (\bibinfo{year}{2003}).

\bibitem[{\citenamefont{Rembieli\'nski and Smoli\'nski}(2002)}]{RS2002}
\bibinfo{author}{\bibfnamefont{J.}~\bibnamefont{Rembieli\'nski}}
  \bibnamefont{and} \bibinfo{author}{\bibfnamefont{K.~A.}
  \bibnamefont{Smoli\'nski}}, \bibinfo{journal}{Phys. Rev. A}
  \textbf{\bibinfo{volume}{66}}, \bibinfo{pages}{052114}
  (\bibinfo{year}{2002}).

\bibitem[{\citenamefont{Soo and Lin}(2004)}]{SL2004}
\bibinfo{author}{\bibfnamefont{C.}~\bibnamefont{Soo}} \bibnamefont{and}
  \bibinfo{author}{\bibfnamefont{C.~C.~Y.} \bibnamefont{Lin}},
  \bibinfo{journal}{Int. J. Q. Inf.} \textbf{\bibinfo{volume}{2}},
  \bibinfo{pages}{183} (\bibinfo{year}{2004}).

\bibitem[{\citenamefont{Terashima and Ueda}(2003{\natexlab{a}})}]{TU2003_1}
\bibinfo{author}{\bibfnamefont{H.}~\bibnamefont{Terashima}} \bibnamefont{and}
  \bibinfo{author}{\bibfnamefont{M.}~\bibnamefont{Ueda}},
  \bibinfo{journal}{Int. J. Q. Inf.} \textbf{\bibinfo{volume}{1}},
  \bibinfo{pages}{93} (\bibinfo{year}{2003}{\natexlab{a}}).

\bibitem[{\citenamefont{Terashima and Ueda}(2003{\natexlab{b}})}]{TU2003_2}
\bibinfo{author}{\bibfnamefont{H.}~\bibnamefont{Terashima}} \bibnamefont{and}
  \bibinfo{author}{\bibfnamefont{M.}~\bibnamefont{Ueda}}, \bibinfo{journal}{Q.
  Inf. and Comput.} \textbf{\bibinfo{volume}{3}}, \bibinfo{pages}{224}
  (\bibinfo{year}{2003}{\natexlab{b}}).

\bibitem[{\citenamefont{Terno}(2003)}]{Terno2003}
\bibinfo{author}{\bibfnamefont{D.~R.} \bibnamefont{Terno}},
  \bibinfo{journal}{Phys. Rev. A} \textbf{\bibinfo{volume}{67}},
  \bibinfo{pages}{014102} (\bibinfo{year}{2003}).

\bibitem[{\citenamefont{Terno}(2005)}]{Terno2005}
\bibinfo{author}{\bibfnamefont{D.~R.} \bibnamefont{Terno}}
  (\bibinfo{year}{2005}), \eprint{arXiv: quant-ph/0508049}.

\bibitem[{\citenamefont{Timpson and Brown}(2002)}]{TB2002}
\bibinfo{author}{\bibfnamefont{C.~G.} \bibnamefont{Timpson}} \bibnamefont{and}
  \bibinfo{author}{\bibfnamefont{H.~R.} \bibnamefont{Brown}}, in
  \emph{\bibinfo{booktitle}{Understending {P}hysical {K}nowlwdge}}, edited by
  \bibinfo{editor}{\bibfnamefont{R.}~\bibnamefont{Lupacchini}}
  \bibnamefont{and} \bibinfo{editor}{\bibfnamefont{V.}~\bibnamefont{Fano}}
  (\bibinfo{publisher}{Department of Philosophy, University of Bologna, CLUEB},
  \bibinfo{year}{2002}).

\bibitem[{\citenamefont{You et~al.}(2004)\citenamefont{You, Wang, Young, Niu,
  Ma, and Xu}}]{YWYNMX2004}
\bibinfo{author}{\bibfnamefont{H.}~\bibnamefont{You}},
  \bibinfo{author}{\bibfnamefont{A.~M.} \bibnamefont{Wang}},
  \bibinfo{author}{\bibfnamefont{X.}~\bibnamefont{Young}},
  \bibinfo{author}{\bibfnamefont{W.}~\bibnamefont{Niu}},
  \bibinfo{author}{\bibfnamefont{X.}~\bibnamefont{Ma}}, \bibnamefont{and}
  \bibinfo{author}{\bibfnamefont{F.}~\bibnamefont{Xu}}, \bibinfo{journal}{Phys.
  Lett. A} \textbf{\bibinfo{volume}{333}}, \bibinfo{pages}{389}
  (\bibinfo{year}{2004}).

\bibitem[{\citenamefont{Zbinden et~al.}(2001)\citenamefont{Zbinden, Brendel,
  Gisin, and Tittel}}]{ZBGT2001}
\bibinfo{author}{\bibfnamefont{H.}~\bibnamefont{Zbinden}},
  \bibinfo{author}{\bibfnamefont{J.}~\bibnamefont{Brendel}},
  \bibinfo{author}{\bibfnamefont{N.}~\bibnamefont{Gisin}}, \bibnamefont{and}
  \bibinfo{author}{\bibfnamefont{W.}~\bibnamefont{Tittel}},
  \bibinfo{journal}{Phys. Rev. A} \textbf{\bibinfo{volume}{63}},
  \bibinfo{pages}{022111} (\bibinfo{year}{2001}).

\bibitem[{\citenamefont{Jordan et~al.}(2006)\citenamefont{Jordan, Shaji, and
  Sudarshan}}]{JSS2006}
\bibinfo{author}{\bibfnamefont{T.~F.} \bibnamefont{Jordan}},
  \bibinfo{author}{\bibfnamefont{A.}~\bibnamefont{Shaji}}, \bibnamefont{and}
  \bibinfo{author}{\bibfnamefont{E.~C.~G.} \bibnamefont{Sudarshan}},
  \emph{\bibinfo{title}{{L}orentz transformations that entangle spins and
  entangle momenta}} (\bibinfo{year}{2006}), \eprint{arXiv: quant-ph/0608061}.

\bibitem[{\citenamefont{Kim and Son}(2005)}]{KS2005}
\bibinfo{author}{\bibfnamefont{W.~T.} \bibnamefont{Kim}} \bibnamefont{and}
  \bibinfo{author}{\bibfnamefont{E.~J.} \bibnamefont{Son}},
  \bibinfo{journal}{Phys. Rev. A} \textbf{\bibinfo{volume}{71}},
  \bibinfo{pages}{014102} (\bibinfo{year}{2005}).

\bibitem[{\citenamefont{Lamehi-Rachti and Mittig}(1976)}]{cab_LRM1976}
\bibinfo{author}{\bibfnamefont{M.}~\bibnamefont{Lamehi-Rachti}}
  \bibnamefont{and} \bibinfo{author}{\bibfnamefont{W.}~\bibnamefont{Mittig}},
  \bibinfo{journal}{Phys. Rev. D} \textbf{\bibinfo{volume}{14}},
  \bibinfo{pages}{2543} (\bibinfo{year}{1976}).

\bibitem[{\citenamefont{Eidelman et~al.}(2004)\citenamefont{Eidelman, Hayes,
  Olive, Aguilar-Benitez, Amsler, Asner, and at~al}}]{cab_PDG2004}
\bibinfo{author}{\bibfnamefont{S.}~\bibnamefont{Eidelman}},
  \bibinfo{author}{\bibfnamefont{K.~G.} \bibnamefont{Hayes}},
  \bibinfo{author}{\bibfnamefont{K.~A.} \bibnamefont{Olive}},
  \bibinfo{author}{\bibfnamefont{M.}~\bibnamefont{Aguilar-Benitez}},
  \bibinfo{author}{\bibfnamefont{C.}~\bibnamefont{Amsler}},
  \bibinfo{author}{\bibfnamefont{D.}~\bibnamefont{Asner}}, \bibnamefont{and}
  \bibinfo{author}{\bibnamefont{at~al}}, \bibinfo{journal}{Phys. Lett. B}
  \textbf{\bibinfo{volume}{592}}, \bibinfo{pages}{1} (\bibinfo{year}{2004}).

\bibitem[{\citenamefont{Bogolubov et~al.}(1975)\citenamefont{Bogolubov,
  Logunov, and Todorov}}]{cab_BLT1969}
\bibinfo{author}{\bibfnamefont{N.~N.} \bibnamefont{Bogolubov}},
  \bibinfo{author}{\bibfnamefont{A.~A.} \bibnamefont{Logunov}},
  \bibnamefont{and} \bibinfo{author}{\bibfnamefont{I.~T.}
  \bibnamefont{Todorov}}, \emph{\bibinfo{title}{Introduction to Axiomatic
  Quantum Field Theory}} (\bibinfo{publisher}{W. A. Benjamin},
  \bibinfo{address}{Reading, Mass.}, \bibinfo{year}{1975}).

\bibitem[{cab()}]{cab_SLAC}
\urlprefix\url{http://www-sld.slac.stanford.edu/sldwww/pubs.html}.

\bibitem[{\citenamefont{Ciborowski}()}]{cab_Ciborowski2006}
\bibinfo{author}{\bibfnamefont{J.}~\bibnamefont{Ciborowski}},
  \bibinfo{howpublished}{private communication}.

\bibitem[{\citenamefont{Barut and R\c{a}czka}(1977)}]{cab_BR1977}
\bibinfo{author}{\bibfnamefont{A.~O.} \bibnamefont{Barut}} \bibnamefont{and}
  \bibinfo{author}{\bibfnamefont{R.}~\bibnamefont{R\c{a}czka}},
  \emph{\bibinfo{title}{Theory of Group Representations and Applications}}
  (\bibinfo{publisher}{PWN}, \bibinfo{address}{Warszawa},
  \bibinfo{year}{1977}).

\bibitem[{\citenamefont{Caban et~al.}(2003)\citenamefont{Caban, Rembieli\'nski,
  Smoli\'nski, and Walczak}}]{cab_CRSW2003_1}
\bibinfo{author}{\bibfnamefont{P.}~\bibnamefont{Caban}},
  \bibinfo{author}{\bibfnamefont{J.}~\bibnamefont{Rembieli\'nski}},
  \bibinfo{author}{\bibfnamefont{K.~A.} \bibnamefont{Smoli\'nski}},
  \bibnamefont{and} \bibinfo{author}{\bibfnamefont{Z.}~\bibnamefont{Walczak}},
  \bibinfo{journal}{Phys.\ Rev.\ A} \textbf{\bibinfo{volume}{67}},
  \bibinfo{pages}{012109} (\bibinfo{year}{2003}).

\bibitem[{\citenamefont{Caban and Rembieli\'nski}(1999)}]{cab_CR1999}
\bibinfo{author}{\bibfnamefont{P.}~\bibnamefont{Caban}} \bibnamefont{and}
  \bibinfo{author}{\bibfnamefont{J.}~\bibnamefont{Rembieli\'nski}},
  \bibinfo{journal}{Phys.\ Rev.\ A} \textbf{\bibinfo{volume}{59}},
  \bibinfo{pages}{4187} (\bibinfo{year}{1999}).

\end{thebibliography}

\end{document}